\documentclass[twocolumn,showpacs,preprintnumbers]{revtex4}
\usepackage{graphicx}
\usepackage{dcolumn}
\usepackage{bm}

\input{tcilatex}

\begin{document}

\preprint{MPI-Gao}
\title{Patterned nanostructure in AgCo/Pt/MgO(001) thin film }
\author{Zhi-Rong Liu and Huajian Gao}
\affiliation{Max Planck Institute for Metals Research, Heisenbergstrasse 3, D-70569
Stuttgart, Germany}
\author{L. Q. Chen}
\affiliation{Department of Materials Science and Engineering, The Pennsylvania State
University, University Park, PA 16802, USA }
\author{Kyeongjae Cho}
\affiliation{Mechanics and Computation Division, Mechanical Engineering Department,
Stanford University, Stanford, CA 94305, USA }
\date{\today}

\begin{abstract}
The formation of patterned nanostructure in AgCo/Pt/MgO(001) thin film is
simulated by a technique of combining molecular dynamics and phase-field
theory. The dislocation (strain) network existing in Pt/MgO is used as a
template whose pattern is transferred to AgCo phase in spinodal
decomposition, resulting in regular arrays of Co islands that are attracted
by the dislocations. The influence of various factors, such as component
concentration and film thickness, is studied. It is found that the spinodal
decomposition of AgCo in this system is mainly characterized by a
competition between a surface-directed layer structure and the
strain-induced patterned structure, where the patterned Ag-Co structure only
dominates in a small range near the interface (less than 10 atomic layers).
However, if the interlayer diffusion can be minimized by controlling film
growth conditions, it is shown that the patterned structure can be formed
throughout the entire film.
\end{abstract}

\maketitle

\section{Introduction}

With the rapid development of the high-density information storage device,
the control of grain size and grain size distribution in magnetic thin films
will be required for the next generation of information storage technology,
where the drive for decreased media noise and increased storage density is
pushing the grain size below the 10 nm regime.\cite{1,2} Present technology
involves the deposition of Co rich magnetic phases onto seed layers whose
grain size determines the grain size of the magnetic layer through grain to
grain epitaxy. For example, in IBM's 10 Gbit/in.$^2$ demonstration, the
magnetic film with an average grain size of 12 nm is obtained where a Co
phase is deposited onto NiAl seed layers.\cite{2} Other methods to control
grain size and grain size distribution were also explored in the labs. It
was shown recently that monodispersed magnetic nanoparticles can be created
by chemical means and deposited onto a substrate to form a highly regular
array of particles.\cite{2a}

There are several possibilities for introducing controllable length-scales
into the film formation process, one of which is the self-organized growth
on a substrate with preferred nano-pattern. For example, flat nanosized Co
dots form in Au(111) substrate with surface reconstruction,\cite{3} and
ordered arrays of vertical magnetic Co pillars can be further obtained by
alternative deposition of Co and Au layers.\cite{4} When Co film is
deposited on the Pt(111), fcc and hcp phases are arranged in regular
patterns to reduce the strain energy.\cite{5} Recently, Kern et al showed
that the strain-relief dislocation network in substrate can be transferred
through heterogeneous nucleation to regular superlattice of almost
mono-dispersed islands with great feasibility.\cite{6}

In a previous paper,\cite{7} the interaction between the spinodal
decomposition of thin film and the dislocation network of substrate was
briefly discussed with AgCo/Pt/MgO thin film as an example. It was revealed
that Co phases are attracted by dislocations in the Pt/MgO interface,
producing an interesting nanostructure. Such mechanism may provide a
possibility to control the length scale for magnetic recording media layer.
For example, after the patterned nanostructure of AgCo phases is established
in spinodal decomposition as a seed layer, a Co-rich conventional medial
alloy can grow on it. In this article, detailed simulations are conducted in
AgCo/Pt/MgO(001) thin film to investigate the formation of nanostructure
under various conditions.

% A major challenge in the nanotechnolgy is to fabricate patterned nanostructures.
% template # nanostructure # magnetic storage # thin film # AgCo spinodal # AgCo/Pt/MgO

% fig.1
\begin{figure}[b]
\includegraphics[width=8.5cm]{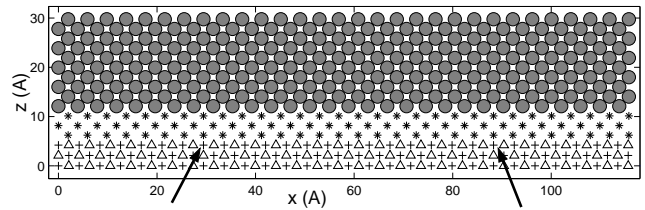}% Here is how to import EPS art
\caption{Schematic graphics of the AgCo/Pt/MgO(001) thin film. The symbols
cross ($+$), triangle ($\triangle$) and stars ($*$) represent atoms Mg, O,
and Pt, respectively, while filled circle ($\bullet$) for AgCo atoms. Pt
atoms trend to sit atop the surface O atoms due to their favorable
interaction energy. The arrows indicate the locations where dislocation will
be generated. Axes are plotted in units of \AA . }
\label{fig01}
\end{figure}

\section{Theoretical model and simulation method}

\subsection{System model}

The system considered here is the AgCo/Pt/MgO(001) thin film (see Fig.~1).
MgO is preferred substrate material in many of the new information storage
and processing devices made from thin metal films.\cite{8} MgO has a f.c.c.
lattice structure with lattice parameter $a_{\mathrm{MgO}}=4.21$\AA , which
is slightly larger than that of Pt ($a_{\mathrm{Pt}}=3.924$\AA ). When Pt
film is deposited on MgO substrate, due to the interactions between film and
substrate, Pt phase will be subject to elastic mismatch strain. The mismatch
strain energy is partly released by the formation of interfacial dislocation
network.\cite{9} The dislocation network of Pt/MgO(001) with cube-on-cube
orientation relationship appears as a square net with dislocation spacing
4.05 nm. Such dislocation network results in periodic variations of strain
in Pt thin film (see Fig.~2). It has been pointed out by theoretical analysis%
\cite{10} and simulation\cite{7} that such periodic strain provides a
nanopattern for the spinodal of a second film on it. In this article, we
consider a AgCo thin film on the strained Pt/MgO(001) (Fig.~1) and simulate
the spinodal decomposition of AgCo phase. As most simulations on spinodal
decomposition, the atoms of the initial AgCo phase are randomly arranged in
regular (f.c.c.) crystal lattice. The model assumes that AgCo is deposited
on Pt/MgO in a disordered state and we calculate the formation of
nanostructure by a subsequent annealing process.

% fig.2
\begin{figure}[tbp]
\includegraphics[width=8.5cm]{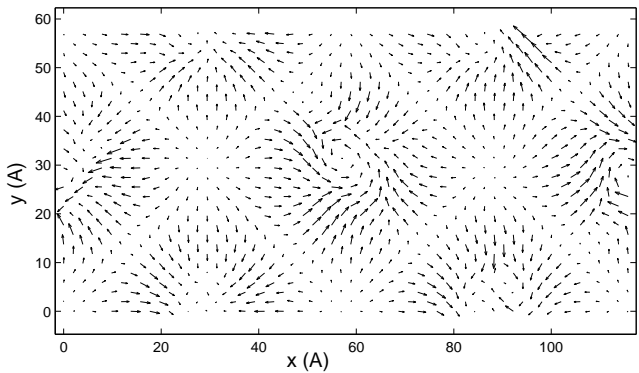}% Here is how to import EPS art
\caption{The strain effect of the as-grown Pt/MgO(001) thin film that is
represented by non-uniform atomic displacements on the surface. A system
with 3 ML of Pt is simulated and the displacements of surface atoms are
plotted. }
\label{fig02}
\end{figure}

% figure 01: schematic graphics of AgCo/Pt/MgO (10*3*3), ideal position, arrow to indicate dislocation
%   SamLiu90  size 10, *o+^
% figure 02: strain of Pt/MgO, 3 ML of Pt,  SamLiu75
%  mi=-0.08, ma=0.05

\subsection{Simulation method}

There are a number of approaches proposed to model the elastic effect on
precipitate morphology in coherent systems,\cite{10a} where the linear
elasticity theory is employed to calculate the elastic energy. In this
article, we adopt an alternative method\cite{7,10b} of combining atomistic
calculations and microscopic mean-field theory, which is applicable for
noncoherent cases, to simulate the effect of dislocation, strain response
and spinodal decomposition in AgCo/Pt/MgO(001) thin film. In this study, the
morphology of the system is described by the position and occupation
probability (atomic concentration) of all atoms, \{$\mathbf{r}_{i},n_{i}$\}.
\{$\mathbf{r}_{i}$\} are used to simulate local atomic motion such as strain
response or dislocation, while \{$n_{i}$\} are used to simulate atomic
diffusion of phase transformation. The atomic concentrations of MgO and Pt
phases are fixed in simulation, while the AgCo phase undergoes a process of
spinodal decomposition. After expressing the empirical potential and free
energy as functions of \{$\mathbf{r}_{i},n_{i}$\} under mean-field
approximation, i.e., $V(\{\mathbf{r}_{i},n_{i}\})$ and $F(\{\mathbf{r}%
_{i},n_{i}\})$, the evolution of atomic positions is given in molecular
dynamics (MD) as\cite{10c} 
\begin{equation}
m_{i}\frac{d^{2}\mathbf{r}_{i}}{dt^{2}}=-\nabla _{i}V(\{\mathbf{r}%
_{i},n_{i}\}),
\end{equation}%
while the diffusion process is described in microscopic mean-field theory as%
\cite{10d,10e} 
\begin{equation}
\frac{dn_{i}}{dt}=\sum_{j}L_{ij}\frac{\partial F(\{\mathbf{r}_{i},n_{i}\})}{%
\partial n_{j}},
\end{equation}%
where $L_{ij}$ is the kinetic coefficient proportional to the inverse
average time of elementary diffusional jumps from site $i$ to $j$. The
timescale of diffusion is mainly determined by the magnitude of $L_{ij}$. In
our calculation, $L_{ij}$ is assigned as a constant $L$ when $|\mathbf{r}%
_{i}-\mathbf{r}_{j}|<R_{\mathrm{diff}}$ and 0 otherwise. Since diffusion is
much slower than elastic response, the system is set in elastic equilibrium
during the diffusion process. In numerical scheme, the simulation loop is as
following: \newline
(1) run $m$ steps of MD by Eq.~(1) with time step $\delta t$ to reach
elastic equilibrium; \newline
(2) run one step of diffusion by Eq.~(2) with time step $dt$; \newline
(3) goto (1). \newline
It is a multiscale scheme overcoming the weakness of small time scale in MD
since $m\delta t\ll dt$. Although the present work employed a microscopic
diffusion equation for a binary system and assumed an atomic exchange
mechanism for diffusion, it is rather straightforward to formulate the
microscopic diffusion using a vacancy mechanism by introducing vacancy as
the third component. With vacancy as the third component, the model can
automatically model the surface morphology changes as a result of surface
energy anisotropy and the balance of surface energies and the interphase
boundary energy between the two phases in the film.\cite{10f} However, as
the first attempt, we assumed the simple exchange mechanism. The vacancy
mechanism will be considered in a future publication.

\subsection{Empirical potential}

The empirical potential between Ag, Co and Pt atoms adopted in current
simulation is the Tight-binding second-moment-approximation (TB-SMA) scheme,%
\cite{11} where the potential is comprised of a binding part and a repulsive
part: 
\begin{equation}
V=\sum_{i}(E_R^i+E_B^i).
\end{equation}
For AgCo phase, to include the influence of atomic concentration, the
following formula is used under mean-field approximation for the potential
(A and B denotes two atomic species, Ag and Co, respectively, and $n_i$, $n_j
$ are the concentrations of A atoms): 
\begin{widetext}
\begin{equation}
 E_R^i=\sum_{j} \left\{
\begin{array}{l}
n_i n_j A_{\rm AA}\exp[-p_{\rm AA} (r_{ij}/r_0^{\rm AA}-1)]  \\
+ \left[ n_{i}(1-n_{j})+(1-n_{i})n_{j} \right]A_{\rm AB}\exp[-p_{\rm AB} (r_{ij}/r_0^{\rm AB}-1)] \\
 + (1-n_i) (1-n_j) A_{\rm BB}\exp[-p_{\rm BB} (r_{ij}/r_0^{\rm BB}-1)]
\end{array}
\right\},
\end{equation}
\begin{equation}
E_B^i=-\left\{ \sum_{j} \left[
\begin{array}{l}
n_i n_j \xi_{\rm AA}\exp[-q_{\rm AA} (r_{ij}/r_0^{\rm AA}-1)]  \\
 +  \left[ n_{i}(1-n_{j})+(1-n_{i})n_{j} \right]\xi_{\rm AB}\exp[-q_{\rm AB} (r_{ij}/r_0^{\rm AB}-1)] \\
 +  (1-n_i) (1-n_j) \xi_{\rm BB}\exp[-q_{\rm BB} (r_{ij}/r_0^{\rm BB}-1)]
\end{array}
\right]^2 \right\}^{1/2},
\end{equation}
\end{widetext}
where $r_{ij}$ represents the distance between atoms $i$ and $j$ and $%
r_{0}^{\alpha\beta}$ is the first-neighbors distance in the $\alpha\beta$
lattice. $A$, $p$, $\xi$, $q$ are some free parameters of the SMA scheme
that are fitted to desired properties of the system. The parameters for the
pure species are available in the literature.\cite{11} For the cross
parameters of Ag-Co interaction, we fit $A$, $\xi$ to the heat of formation
(+26 kJ/g atom)\cite{12} and lattice parameter ($3.81$ $\mathrm{\mathring{A}}
$ that is assumed to be the average value between pure Ag and Co) of the Ag$%
_{0.5}$Co$_{0.5}$ disordered phase, while $p$ and $q$ are simply assigned to
the average values of pure species. The result obtained is shown in Table I.
Based on the potential scheme Eqns.~(3-5) and the fitted parameters, the
formation heat and the lattice constant at any atomic concentration can be
calculated, which are depicted as Fig.~3. The curve of formation heat can be
described by the parabolic law very well. The lattice constant approximately
obeys the Vegard's law. The elastic constants are also calculated (see
Fig.~4). It appears that the elastic constants monotonously change with
concentration. However, it should be noted that the curves do not satisfy
the linear law that is assumed by many spinodal decomposition simulations.

% fig.3
\begin{figure}[b]
\includegraphics[width=8.5cm]{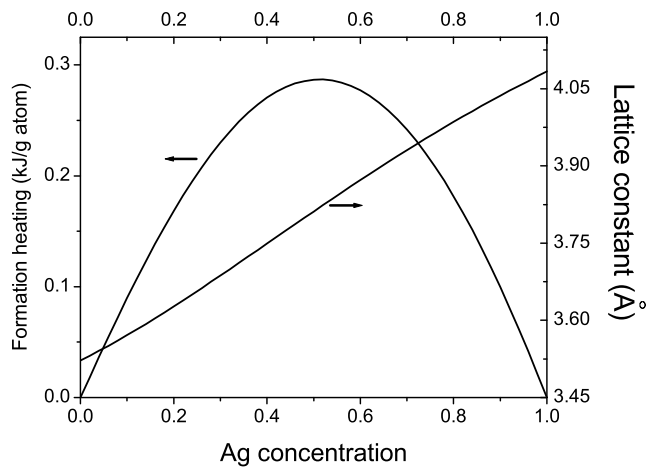}% Here is how to import EPS art
\caption{Heat of formation and lattice constant of disordered AgCo phase as
functions of Ag concentration. The curves are calculated by using the
mean-field TB-SMA potential [Eqns.~(3-5)] with parameters in Table I. }
\label{fig03}
\end{figure}

% fig.4
\begin{figure}[b]
\includegraphics[width=8.5cm]{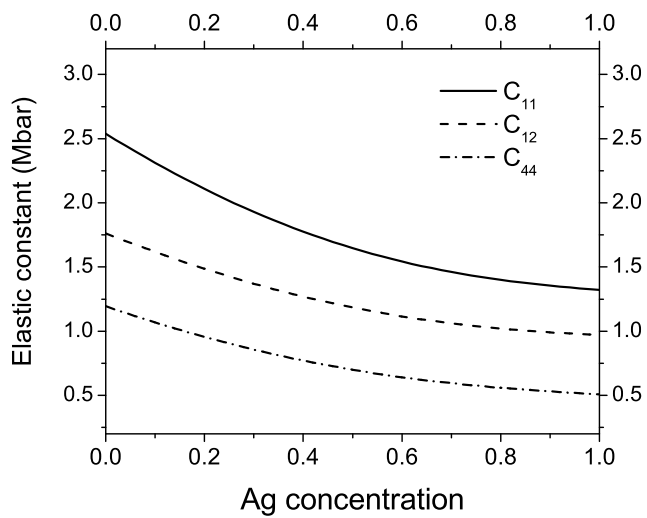}% Here is how to import EPS art
\caption{Elastic constants of disordered AgCo phase as functions of Ag
concentration under the mean-field TB-SMA potential. }
\label{fig04}
\end{figure}

% figure 03: crystal lattice, mixing energy
% figure 04: elastic constant
% dmd26

The cross parameters of Ag-Pt and Co-Pt are also listed in Table I, which
are fitted in a similar process. The heat of formation is assumed to be zero.

The importance of the Pt/MgO substrate in the current simulation lies in its
dislocation network and the corresponding strain effect. We keep the MgO
atoms fixed in simulation and describe the interactions between Pt and MgO
as the Lenard-Jones potential: 
\begin{equation}
V_{\mathrm{LJ}}=4\varepsilon\left[ \left(\frac{\sigma}{r}\right)^{12}-\left(%
\frac{\sigma}{r}\right)^{6} \right].
\end{equation}
The parameters are adopted as: $\varepsilon_{\mathrm{Mg-Pt}}=0.336$eV, $%
\sigma_{\mathrm{Mg-Pt}}=2.3$\AA , $\varepsilon_{\mathrm{O-Pt}}=1.34$eV, $%
\sigma_{\mathrm{O-Pt}}=1.783$\AA , with a distance cutoff $3.4$\AA . With
these parameters, isolated Pt atoms can be adsorbed by MgO substrate by
sitting $1.99\mathrm{\mathring{A}}$ atop the surface O atoms of MgO(001)
with an adsorption energy of 2.35 eV as indicated by density-functional
calculations.\cite{13} MD simulation shows that such potential can produce
dislocation network as required and the corresponding periodic strain (see
Fig.~2).

\begin{table}[t]
\caption{TB-SMA parameters for Ag-Co-Pt }
\label{tab:table1}% \begin{ruledtabular}
\begin{ruledtabular}
\begin{tabular}{lccccr}
 & $A$ (eV) & $\xi$ (eV) & $p$ & $q$ & $r_0$ (\AA) \\
\hline
Ag-Co & 0.0752 & 1.1944 & 11.261 & 2.7127 & 2.696  \\
Ag-Pt & 0.1988 & 1.9399 & 10.770 & 3.5711 & 2.832 \\
Co-Pt & 0.1449 & 1.9959 & 11.108 & 3.145 & 2.667 \\
\end{tabular}
\end{ruledtabular}
% \end{ruledtabular}
\end{table}

% fig.5
\begin{figure}[tbp]
\includegraphics[width=8.5cm]{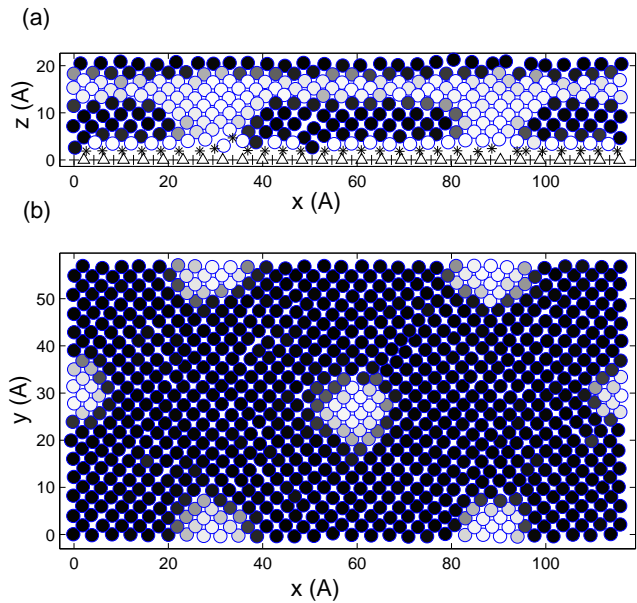}% Here is how to import EPS art
\caption{Atomic configuration of the system with 1 ML of Pt and 10 ML of
AgCo ($c_{\mathrm{Ag}}=0.5$) at $t^{*}=70$. $240000$ molecular dynamics
steps are used. The symbols $+$, $\triangle$ and $*$ represent Mg, O, and Pt
atoms, respectively. The occupation probability $n_i$ of AgCo atoms is
visualized by the filling darkness of the circle at $\mathbf{r}_i$, with
white representing low value (Co) and black representing high value (Ag).
(a) The $x$-$z$ monolayer located at $y=0$. (b) The third $x$-$y$ monolayer
of AgCo phase (counting from the bottom to the surface). }
\label{fig05}
\end{figure}

\section{Results and Discussions}

\subsection{Basic feature}

A 3D system of AgCo/Pt/MgO(001) thin film with 1 monolayer (ML) of Pt and 10
ML of AgCo phase is simulated. Every monolayer (in $xy$-plane) contains $%
30\times 30$ AgCo atoms and periodic boundary conditions are applied in $x$-
and $y$-axes. Free boundary condition is used in $z$-axis. The average Ag
concentration is $c_{\mathrm{Ag}}=0.5$. The diffusion cutoff in one
diffusion step is $R_{\mathrm{diff}}=3.53$\AA , i.e., $L_{ij}=L$ for the
nearest neighbors and $L_{ij}=0$ in other cases. A reduced time is defined
as $t^{*}=t/L$ to measure the simulation process. The time step of MD is
kept as $\delta t=0.15\times 10^{-15} s$, and the temperature is 600 K. The
time step of diffusion simulation is dynamically adjusted to ensure the
convergence of the result.

The simulation result corresponding to $t^{*}=70$ is depicted in Fig.~5. $%
240000$ MD steps are conducted in simulation. We can see that the surface of
thin film is covered by Ag phase. The reason is that the surface energy of
Ag ($1.3$ J/m$^2$) is half that of Co ($2.71$ J/m$^2$).\cite{14} Similar
case occurs in CuCo system where Cu phase will diffuse onto the surface to
form a metastable alloy in epitaxial growth.\cite{15} Due to the influence
of surface, AgCo decomposed into layer structure near the surface. In fact,
it is a kind of surface-directed spinodal decomposition.\cite{16}
Interesting phenomenon is observed near the AgCo/Pt interface: Co atoms are
attracted to the strained regions near the dislocations because their atomic
radius are smaller than Ag. As a result, AgCo decomposed phase produces
interesting 2D nanostructure pattern [Fig.~5(b)]. So the basic feature of
spinodal decomposition in AgCo/Pt/MgO(001) system can be described as a
competition between the surface-directed layer structure and the
substrate-dislocation-induced nanopattern. These features are consistent
with previous results in 2D approximation.\cite{7}

% fig. 05: SamLiu47a-1, SamLiu47a-2. t*=70, N=240000, Rdiff=3.53A, dt=0.015
% j0=4
% subplot('position',[0.2 0.75 0.6 0.18]);  subplot('position',[0.2 0.1 0.6 0.47]); size 8

% fig.6
\begin{figure}[tbp]
\includegraphics[width=8.5cm]{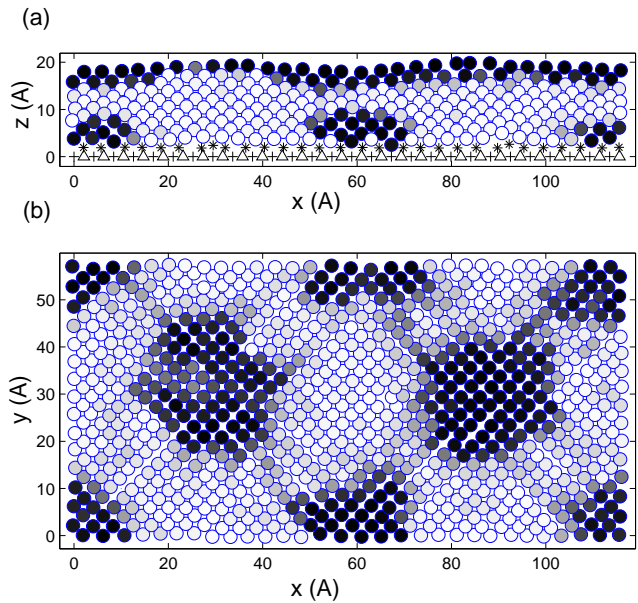}% Here is how to import EPS art
\caption{Snap-shot of phase morphologies of the system with 1 ML of Pt and
10 ML of AgCo ($c_{\mathrm{Ag}}=0.25$). (a) The $x$-$z$ monolayer located at 
$y=0$. (b) The second $x$-$y$ monolayer of AgCo phase. }
\label{fig06}
\end{figure}

\subsection{Effect of AgCo concentration}

In this section, we investigate the effect of AgCo concentration on the
spinodal decomposition. Two Ag concentrations, $c_{\mathrm{Ag}}=0.25$ and $%
0.75$, are considered, and the results are compared with that of $c_{\mathrm{%
Ag}}=0.5$ in Section III.A.

A result of $c_{\mathrm{Ag}}=0.25$ is given in Fig.~6. The layer structure
near the surface is similar to the case of $c_{\mathrm{Ag}}=0.5$. The
difference occurs in the nanostructure near the interface: due to the
reduction of Ag concentration, Ag-rich regions near the interface shrink in
volume, and the nano-pattern changes from Co islands in Ag matrix
[Fig.~5(b)] into Ag islands in Co matrix [Fig.~6(b)]. When the system is
evolved in longer time, it is observed in simulation that the flat Ag
islands get smaller and smaller. It means the surface Ag phase tends to
absorb the Ag islands near the interface.

% fig. 06: SamLiu47e-3, SamLiu47e-4, n=30, j0=3

% fig.7
\begin{figure}[tbp]
\includegraphics[width=8.5cm]{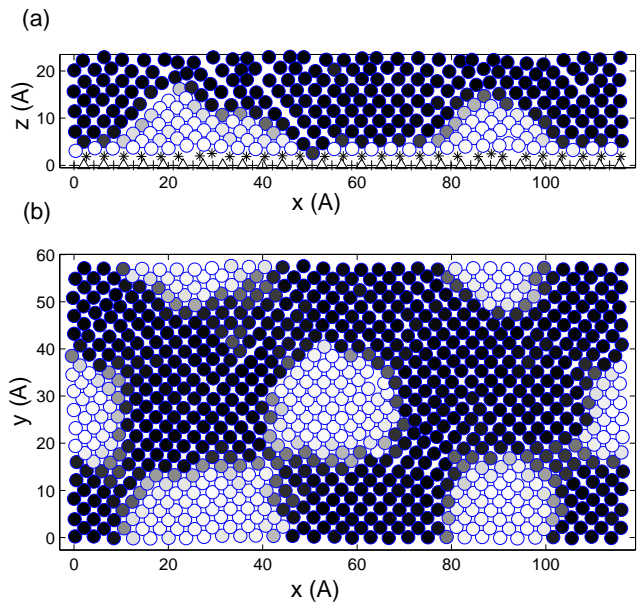}% Here is how to import EPS art
\caption{Atomic configuration of the system with 1 ML of Pt and 10 ML of
AgCo ($c_{\mathrm{Ag}}=0.75$). (a) The $x$-$z$ monolayer located at $y=0$.
(b) The third $x$-$y$ monolayer of AgCo phase. }
\label{fig07}
\end{figure}

Now, we come to the case of $c_{\mathrm{Ag}}=0.75$ (Fig.~7). Compared with
the result of $c_{\mathrm{Ag}}=0.5$, there is no layer structure of Co phase
near the surface in present case. All Co atoms condense as cone-like islands
on the dislocations of the substrate. They are covered by Ag-rich phase that
spread to the surface. When $c_{\mathrm{Ag}}=0.5$, the Co phases near the
interface also assume the shape of cone-like islands [Fig.~5(a)], but the
cones stand upside down, merging with the layer structure near the surface.

It is noted that the Ag phase is important in forming the Co islands in the
above process. If there is no Ag phase, Co thin film will grow in layer
structure at small thickness. Ag phase acts as equivalent surfactant in
assisting Co to decompose into islands. The present study suggests it may be
possible to produce nanostructure by the following process. First, Co
islands can be formed via spinodal decomposition of AgCo on periodic
strained substrate. Subsequently, the Ag phase can be removed by chemical
etching, leaving a regular array of Co islands. Experiments will be needed
to test this conjecture.

% fig. 07: SamLiu47d-1, SamLiu47d-2, j0=4

% fig.8
\begin{figure}[tbp]
\includegraphics[width=8.5cm]{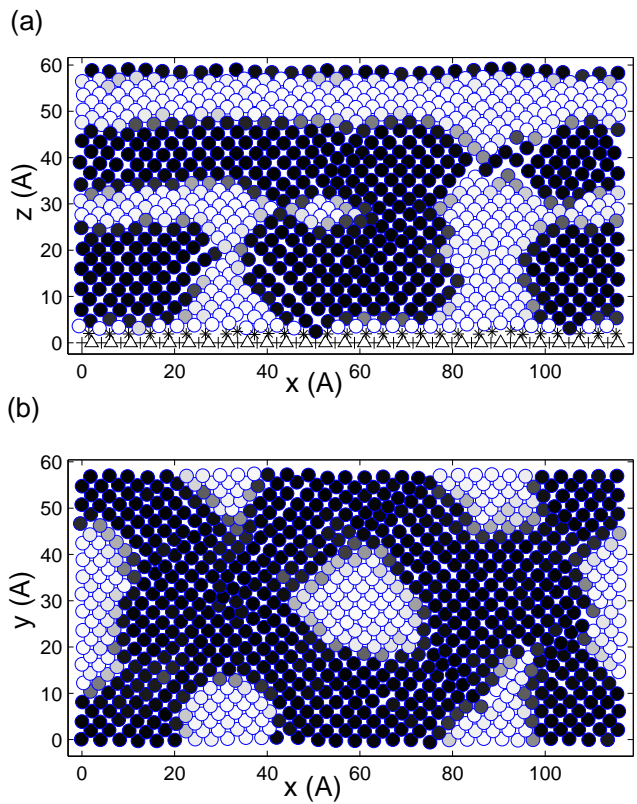}% Here is how to import EPS art
\caption{Atomic configuration of the system with 1 ML of Pt and 30 ML of
AgCo ($c_{\mathrm{Ag}}=0.5$). (a) The $x$-$z$ monolayer located at $y=0$.
(b) The third $x$-$y$ monolayer of AgCo phase. }
\label{fig08}
\end{figure}

\subsection{Effect of AgCo phase thickness}

To explore the effect of AgCo phase thickness, a system with 30 ML of AgCo
with $c_{\mathrm{Ag}}=0.5$ is simulated with the result in Fig.~8. The basic
characteristics is similar to the above results: a competition between the
layer structure near the surface and the patterned structure near the
interface. The patterned structure induced by the dislocations of the
substrate decays within a distance of about ten monolayers while the layer
structure presides in most regions. Surface energy is on the same order as
the atomic cohesive energy while strain energy is higher order variation of
the cohesive energy. So the surface effect dominates at the current film
thickness.

In order to quantitatively analyze the competition between two kinds of
structures, we define the pattern correlation function of two layers as: 
\begin{equation}
R_{mn}=\frac{\mathbf{V}_{m}\cdot\mathbf{V}_{n}} {|\mathbf{V}_{m}|\cdot|%
\mathbf{V}_{n}|}-R_{mn}^0,
\end{equation}
where $\mathbf{V}_{m}$, $\mathbf{V}_{n}$ are vectors composed of atomic
concentrations within layers $m$ and $n$. $R_{mn}^0$ is the calculation
result when there is no correlation between two patterns: 
\begin{equation}
R_{mn}^0=\frac{\sum_{\mu}{V}_{m\mu}\cdot \sum_{\nu}{V}_{n\nu}} {N_{\mathrm{%
layer}}|\mathbf{V}_{m}|\cdot|\mathbf{V}_{n}|},
\end{equation}
where ${V}_{m\mu}$ are components of $\mathbf{V}_{m}$ and $N_{\mathrm{layer}}
$ is the number of components.

% fig.9
\begin{figure}[tbp]
\includegraphics[width=8.5cm]{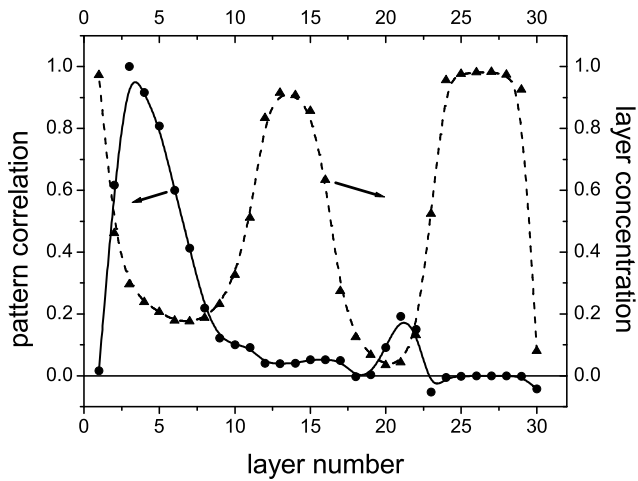}% Here is how to import EPS art
\caption{Pattern correlation function (solid line and circles) and layer Ag
concentration (dashed line and triangles) of AgCo phase in the system with 1
ML of Pt and 30 ML of AgCo ($c_{\mathrm{Ag}}=0.5$). The pattern correlation
function is normalized to make sure the self-correlation is equal to 1 ($%
R_{mm}=1$ for $m=3$ in this case). }
\label{fig09}
\end{figure}

% fig.10
\begin{figure}[tbp]
\includegraphics[width=8.5cm]{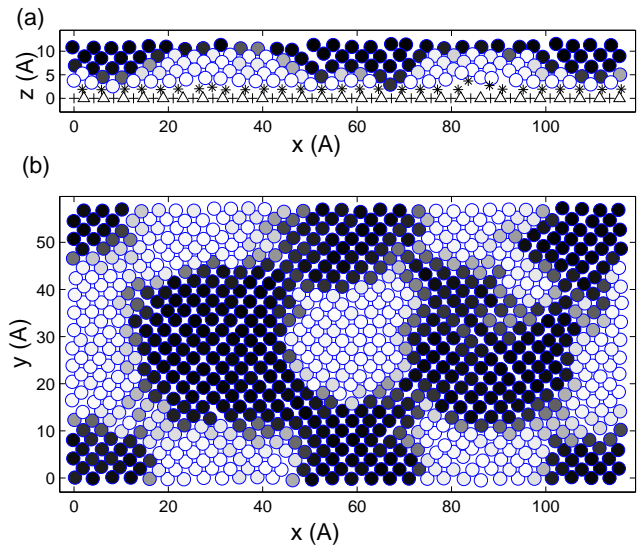}% Here is how to import EPS art
\caption{Atomic configuration of the system with 1 ML of Pt and 5 ML of AgCo
($c_{\mathrm{Ag}}=0.5$). (a) The $x$-$z$ monolayer located at $y=0$. (b) The
third $x$-$y$ monolayer of AgCo phase. }
\label{fig10}
\end{figure}

The third AgCo layer has a representative pattern induced by the dislocation
of substrate. We calculate the correction function of all AgCo layers with
respect to the third one and plot the result in Fig.~9, together with the
layer concentration. It is clearly demonstrated that surface-directed
spinodal decomposition (dashed line) penetrates throughout the thin film
while the pattern structure (solid line) only exists in the range of ten
monolayers near the interface. It suggests that the film thickness should be
limited to small values if one is interested in producing nanostructure by
spinodal decomposition on strained substrate. For example, when the
thickness of AgCo phase decreases to 5 ML, a regular array of Co islands can
be obtained in simulation (see Fig.~10).

% Fig. 08: 1+30, SamLiu70a
% Fig. 09: pattern correlation function with the third one, normalized the largest one to be 1.
% Fig. 10: 1+5, SamLiu72, j0=4

% fig.11
\begin{figure}[tbp]
\includegraphics[width=8.5cm]{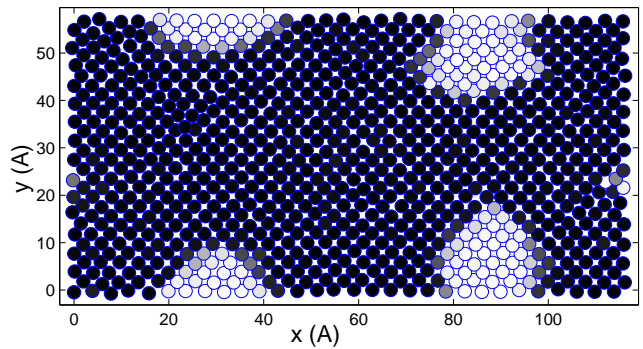}% Here is how to import EPS art
\caption{Atomic configuration of the system with 3 ML of Pt and 30 ML of
AgCo ($c_{\mathrm{Ag}}=0.5$). The third $x$-$y$ monolayer of AgCo phase is
depicted here. }
\label{fig11}
\end{figure}

\subsection{Effect of Pt thickness}

The strain induced by the dislocation declines in Pt phase. If the thickness
of Pt phase is large, the strain acting on AgCo phase will be too weak to
induce nano-pattern.

Fig.~11 displays the simulation results on the system with 3 ML of Pt and 30
ML of AgCo. It can be seen that there are only two Co islands within the
simulation domain, in contrast to four islands in the previous cases. The
other two Co islands, which locate at $y=30\mathrm{\mathring{A}}$ in the
system with 1 ML of Pt, now disappear. It means the strain spreading through
3 ML of Pt is losing its capability of attracting Co atoms to form islands.
3 ML is the critical thickness of Pt phase to produce patterned structure in
the current case.

It has been illuminated in the previous sections that the influence of
surface is competing with that of dislocation strain. With decreasing AgCo
thickness, the structure near interface is more and more affected by the
surface. We conduct simulation on the system with 3 ML of Pt and 10 ML of
AgCo. Co islands appear only in the very early stage, and the final
structure is a complete layer structure without any Co island.

Our analysis suggests there exists a critical value for the periodic strain
of substrate to induce nanostructure in thin film. A previous study\cite{10}
indicates that the imposed strain provides a pre-existing preference
wavelength for the spinodal decomposition in initial stage, regardless of
how small the strain is. The simulation results in this section shows that
such preference wavelength may be lost in later stage and self-organized
nanostructuremay fail to realize if the strain is less than a critical value.

% Fig. 11: SamLiu71a, two Co islands only

% fig.12
\begin{figure}[tbp]
\includegraphics[width=8.5cm]{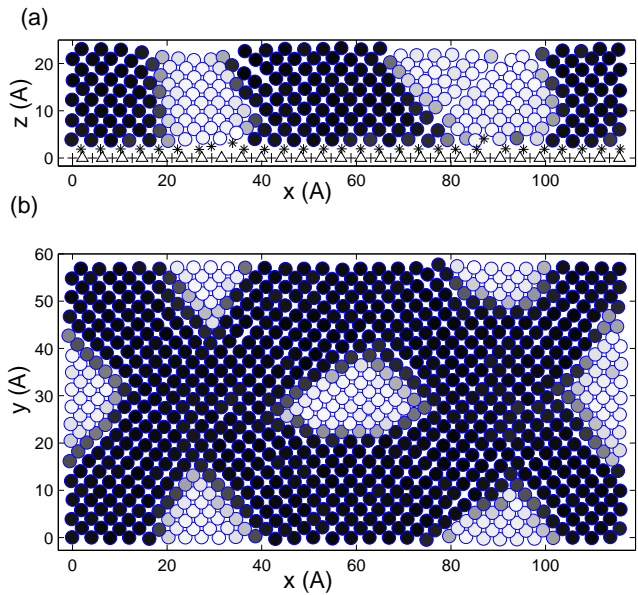}% Here is how to import EPS art
\caption{Atomic configuration of the thin film grown layer by layer where
atomic diffusion is restricted within the same layers. The system has 1 ML
of Pt and 10 ML of AgCo ($c_{\mathrm{Ag}}=0.75$). (a) The $x$-$z$ monolayer
located at $y=0$. (b) The third $x$-$y$ monolayer of AgCo phase. }
\label{fig12}
\end{figure}

\subsection{Film growth: layer by layer}

The above research simulates the annealing process of the disordered AgCo
phase where atomic diffusion occurs in the whole thin film (bulk diffusion).
In experiments, surface diffusion is much faster than bulk diffusion. In the
epitaxial growth of thin film, the system is produced in layer by layer. If
one controls growth conditions carefully to prohibit interlayer diffusion
while fully permitting intralayer diffusion, the resulting nanostructure can
be markedly different (an interesting example in experiment is the formation
of metastable CuCo alloy in thin film\cite{15}). We have simulated such a
growth process by confining the diffusion to within each layer. The result
of a system with 10 ML of AgCo and $c_{\mathrm{Ag}}=0.75$ is plotted in
Fig.~12. No layer structure is observed since interlayer diffusion is
prohibited. The patterned nanostructure near interface penetrates through
the whole film to the surface. This would result in the phenomena of growth
on seed layer, in which case an as-grown AgCo layer influenced by the
substrate strain and shape in nano-pattern acts as seed layer for the
forthcoming growth. Since AgCo is a system with spinodal decomposition, Ag
atom is attracted to Ag atom while Co to Co, so that the nano-pattern of the
as-grown layer is transferred as film grows.

% fig. 12: SamLiu83g

\section{Conclusions}

Based on the above simulation results, it is concluded that patterned
nanostructure can be obtained in AgCo/Pt/MgO thin film. The main feature of
AgCo spinodal decomposition in this system is the competition between the
surface-directed layer structure and the strain-induced patterned structure.
Strain effect is weaker than the surface effect in the current cases. The
patterned Ag-Co structure occurs only within a short range near the
interface (no more than 10 monolayers), and the thickness of Pt phase should
be small enough to effectively transfer the strain of dislocation in Pt/MgO
interface to the AgCo phase to produce nanostructure. However, if spinodal
decomposition of AgCo phase occurs within monolayers by careful film growth
controlling, surface-directed layer structure will be prohibited and the
patterned structure will form in a wide range.

\section*{Acknowledgment}

This work was supported by a Max Planck Post-Doc Fellowship for ZL and by US
National Science Foundation through Grant CMS-0085569. The code of MD
simulation in this work comes from the program GONZO of Stanford Multiscale
Simulation Laboratory. HG also acknowledges support by Yangze lecture
professorship of China.

\end{document}